\documentclass[pre,final,twocolumn,showpacs]{revtex4}
\usepackage{amssymb,amsmath}
\usepackage{bm}
\usepackage{dcolumn}
\usepackage{graphicx}
\begin{document}

\title{Short-time dynamics and critical behavior of three-dimensional site-diluted Ising model}

\author{%%
\firstname{Pavel~V.}~\surname{Prudnikov},
\firstname{Vladimir~V.}~\surname{Prudnikov},
\firstname{Aleksandr~S.}~\surname{Krinitsyn},
\firstname{Andrei~N.}~\surname{Vakilov},
\firstname{Evgenii~A.}~\surname{Pospelov}.
 }%%
\affiliation{%
Department of Theoretical Physics, Omsk State University,  Pr. Mira 55A, Omsk 644077, Russia
}%

\email{prudnikp@univer.omsk.su}

\date{\today}

\begin{abstract}
Monte Carlo simulations of the short-time dynamic behavior are reported
for three-dimensional weakly site-diluted Ising model with spin
concentrations $p=0.95$ and 0.8 at criticality.
In contrast to studies of the critical behavior of the pure systems by
the short-time dynamics method, our investigations of site-diluted Ising model
have revealed three stages of the dynamic evolution characterizing a crossover
phenomenon from the critical behavior typical for the pure systems to behavior
determined by the influence of disorder. The static and dynamic critical
exponents are determined with the use of the corrections to scaling for systems
starting separately from ordered and disordered initial states.
The obtained values of the exponents demonstrate a universal behavior of
weakly site-diluted Ising model in the critical region. The values of the exponents
are compared to results of numerical simulations which have been obtained in various
works and, also, with results of the renormalization-group description of this model.
\end{abstract}


\maketitle

\section{Introduction}

The investigation of critical behavior of disordered systems remains one of the
main problems in condensed-matter physics and excites a great interest, because all
real solids contain structural defects \cite{Stinchcombe,Folk}. The structural disorder
breaks the translational symmetry of the crystal and thus greatly complicates the theoretical
description of the material. The influence of disorder is particularly important near the
critical point where behavior of a system is characterized by anomalous large response on any
even weak perturbation. The description of such systems requires the development of
special analytical and numerical methods.

The effects produced by weak quenched disorder on critical phenomena have been studied
for many years \cite{Harris1,Harris2,Khmelnitskii,Emery,Grinstein1,Grinstein2}. According to
the Harris criterion \cite{Harris1}, the disorder affects the critical behavior only if $\alpha$, the
specific heat exponent of the pure system, is positive. In this case a new universal critical
behavior, with new critical exponents, is established. In contrast, when $\alpha < 0$, the
disorder appears to be irrelevant for the critical behavior. Only systems whose effective
Hamiltonian near the critical point is isomorphic to the Ising model satisfy this criterion.

A large number of publications is devoted to the study of the critical behavior of diluted
Ising-like magnets by the renormalization-group (RG) methods, the numerical Monte Carlo methods,
and experimentally (for a review see Refs.~\cite{Folk,Selke,Cardy}, and \cite{P-V}).
The ideas about replica symmetry breaking in the systems with quenched disorder were
presented in Refs.~\cite{Sherrington,Dotsenko}.
A refined RG analysis of the problem has shown the stability of the critical behavior of
weakly disordered three-dimensional systems with respect to the replica symmetry breaking
effects \cite{PrudnikovRSB}.
All obtained results confirm the existence of a new universal class of the critical behavior,
which is formed by diluted Ising-like systems. However, it remains unclear, whether
the asymptotic values of critical exponents are independent of the rate of dilution of
the system, how the crossover effects change these values, and whether two or more regimes
of the critical behavior exist for weakly and strongly disordered systems.
These questions are the subjects of heated discussions \cite{Folk,Berche}.

For critical dynamic systems, traditionally it is believed that universal scaling behavior
is realized in the long-time regime of dynamic evolution. However, at first in the paper
\cite{Janssen}, it was shown that systems, starting from macroscopic non-equilibrium initial
states, demonstrate a universal scaling behavior on the macroscopic short-time stages of
their dynamic process which is characterized by initial slip exponents $\theta$ and $\theta'$
for the response functions $G(r,t,t')$ and the order parameter $m(t)$ (magnetization for
ferromagnetic systems)
\begin{equation} \label{eq1}
G(r,t,t') \sim (t/t')^{\theta}, \ \ \ m(t) \sim t^{\theta'}.
\end{equation}
A remarkable property of this relaxation process is the increase of magnetization $m(t)$ from
a non-zero initial magnetization $m_0 \ll 1$ at short times
$t < t_{cr} \sim m_0^{-1/(\theta'+\beta/z\nu)}$.
The initial rise of magnetization is changed to the well known decay $m(t) \sim t^{-\beta/z\nu}$
for $t \gg t_{cr}$ \cite{Janssen}.
The critical exponents $\theta$ and $\theta'$ depend on the dynamic universality class \cite{H-H}
and have been calculated by the RG method for a number of dynamic models \cite{CalabGambReview2005}
such as the model with a non-conserved order parameter \cite{Janssen,Prudnikov2008} (model A), the model with an order parameter coupled to a
conserved density \cite{Oerding_93_1} (model C), and the models with reversible mode coupling
\cite{Oerding_93_2} (models E, F, G, and J). The universal scaling behavior of the initial
stage of the critical relaxation for pure systems has been verified by extensive numerical
simulations \cite{Huse,Zheng,Jaster,Luo}. The developed method in these papers of short-time
critical dynamics gives the possibility to determine both the static critical exponents $\nu$,
$\beta$, and dynamic critical exponents $z$ and $\theta'$ in the macroscopic short-time regime
of the critical relaxation. However, a number of publications devoted to the numerical study
of disorder influence on non-equilibrium critical relaxation by the short-time dynamic method
is surprisingly little. We know the papers \cite{Yin,Schehr,SchehrPRE} in which the non-equilibrium critical
dynamics of the three-dimensional (3D) site-diluted Ising ferromagnets with quenched
point-like defects is investigated and the values of the initial slip exponent $\theta'$ \cite{Schehr}
and an exponent $C_a$ for autocorrelation function \cite{SchehrPRE} are
determined for systems with different spin concentrations. The obtained universal value
$\theta'=0.10(2)$ is in good agreement, as it is insisted in \cite{Schehr}, with the RG
estimate for $\theta'= 0.0868$ calculated in the two-loop approximation in Ref.~\cite{Oerding}
with the use of $\varepsilon$-expansion method, where $\varepsilon = 4 - d$, with $d$ is the
spatial dimension. However, some assumptions introduced during investigations
in Ref.~\cite{Schehr} and discussed below precludes from consenting to this value
$\theta'=0.10(2)$ as confirmation of the RG estimate validity.
In our paper \cite{PrudnikovPTP} the integrated Monte Carlo simulations of the short-time
dynamic behavior are reported for 3D Ising and XY models with long-range correlated disorder
at criticality, in the case corresponding to linear defects. Both static and dynamic
critical exponents are determined for systems starting separately from ordered and disordered
initial states. The obtained values of the exponents are in good agreement with results of
the field-theoretic description of the critical behavior of these models in the two-loop
approximation \cite{PrudnikovPR00}.

In the present paper, we numerically investigate the short-time critical dynamics with a
non-conserved order parameter (model A)\cite{H-H} in the 3D site-diluted Ising systems with
spin concentrations $p=0.95$ and 0.8. In the following section, we introduce the 3D Ising
model with quenched point-like defects and scaling relations for the short-time critical
dynamics. In Sec. III, we derive the critical short-time dynamics in Ising systems starting
separately from ordered and disordered initial states. Critical exponents obtained under these
two conditions with the use of the corrections to scaling are compared. The final section
contains analysis of the main results, their comparison to results of other investigations
and our conclusions.

\section{Description of the model and methods}

We have considered the following 3D site-diluted ferromagnetic Ising model Hamiltonian defined
in a cubic lattice of linear size $L$ with periodic boundary conditions:
\begin{eqnarray}
H =-J\sum_{\langle i,j\rangle}p_i p_j S_i S_j,
\end{eqnarray}
where the sum is extended to the nearest neighbors, $J>0$ is the short-range exchange
interaction between spins $S_i$ fixed at the lattice sites and assuming values of $\pm 1$.
Nonmagnetic impurity atoms form empty sites. In this case, occupation numbers $p_i$ assume
the value 0 or 1 and are described by the distribution function
\begin{equation}
P(p_i)= (1-p)\delta (p_i)+p\delta (1-p_i),
\end{equation}
with  $p=1-c$, where  $c$ is the concentration of the impurity atoms.

In this paper we have investigated systems with the spin concentrations $p = 0.95$ and 0.8.
We have considered the cubic lattices with linear size $L = 128$. The Metropolis algorithm
has been used in simulations. We consider only the dynamic evolution of systems described
by the model A in the classification of Hohenberg and Halperin \cite{H-H}. The Metropolis
Monte Carlo scheme of simulation with the dynamics of a single-spin flips reflects the
dynamics of model A and enables us to compare obtained critical exponents
$\theta'$ and $z$ to the results of RG description of the non-equilibrium relaxation
of this model.

According to the argument of Janssen, \textit{et al.} \cite{Janssen} obtained with the
RG method and $\varepsilon$-expansion, one may expect a generalized scaling relation for
the $k$th moment the magnetization
\begin{equation} \label{mk}
m^{(k)} \left( t, \tau, L, m_0 \right)=b^{- k\beta / \nu } m^{(k)} \left( b^{-z} t, \ b^{1/\nu} \tau, \ b^{-1}L, \ b^{x_0}m_0 \right),
\end{equation}
is realized after a time scale $t_{mic}$ which is large enough in a microscopic sense but
still very small in a macroscopic sense. In Eq.~(\ref{mk}), $b$ is a spatial rescaling factor,
$\beta$ and $\nu$ are the well-known static critical exponents, and $z$ is the dynamic
exponent, while the new independent exponent $x_0$ is the scaling dimension of the initial
magnetization $m_0$ and $\tau=(T-T_c)/T_c$ is the reduced temperature.

Since the system is in the early stage of the evolution the correlation length is
still small and finite size problems are nearly absent. Therefore we generally consider
$L$ large enough ($L=128$) and skip this argument. We now choose the scaling factor
$b = t^{1/z}$ so that the main $t$-dependence on the right is cancelled. Applying the
scaling form (\ref{mk}) for $k=1$ to the small quantity $t^{x_0/z}m_0$, one obtains
\begin{eqnarray} \label{theta'}
m(t,\tau,m_0) &\sim& m_0 t^{\theta^{'}} F(t^{1/\nu z}\tau,t^{x_0/z}m_0)  \\
              &=& m_0 t^{\theta^{'}}(1+at^{1/\nu z}\tau)+O(\tau^2, m_0^2), \nonumber
\end{eqnarray}
where $\theta^{'}=(x_0 - \beta/\nu)/z$ has been introduced. It was shown in
Ref.~\cite{Janssen} that the critical exponents $\theta$ and $\theta'$ in
Eq.~(\ref{eq1}) are related by the scaling relation $\theta' = \theta + (2-z-\eta)/z$,
therefore independent exponent is one of them ($\theta$ or $\theta'$).
For $\tau = 0$ and small enough $t$ and $m_0$, the scaling dependence for magnetization
(\ref{theta'}) takes the form $m(t) \sim  t^{\theta'}$. The time scale of a critical initial
increase of the magnetization is $t_{cr} \sim m_0^{-z/x_0}$. However, in the limit
of $m_0 \to 0$, the time scale goes to infinity. Hence the initial condition can leave its
trace even in the long-time regime. For illustration we give in Fig.~1 the time evolution
of the magnetization $m(t)$ from the initial state with magnetization $m_0=0.03$
at $T_c=3.49948$ as a result of Monte Carlo simulation of samples with spin concentration
$p=0.8$ and with linear size $L=128$.

\begin{figure}
\includegraphics[width=0.45\textwidth]{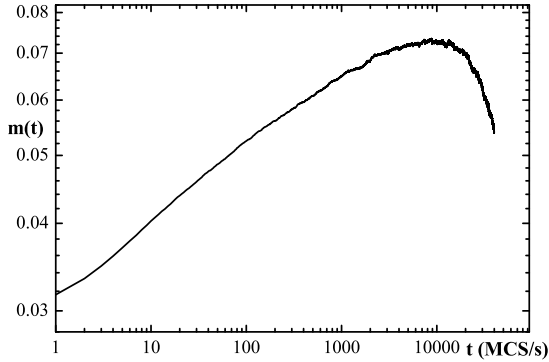}\\
\caption{\label{fig:1}
Time evolution of the magnetization $m(t)$
from the initial state with magnetization $m_0=0.03$ at $T_c=3.49948$ as a result
of Monte Carlo simulation of samples with spin concentration $p=0.8$ and with linear size
$L=128$.\\[5mm]
} \end{figure}

If $\tau \neq 0$, the power law behavior is modified by the scaling function $F(t^{1/z\nu})$
with corrections to the simple power law, which will be dependent on the sign of $\tau$.
Therefore, simulation of the system for temperatures near the critical point allows
to obtain the time-dependent magnetization with nonperfect power behavior and
the critical temperature $T_c$ can be determined by interpolation.

For the site-diluted Ising model, we measured the time evolution of the magnetization
determined as follows:
\begin{equation} \label{m1}
m(t)=\left[ \left<  \frac{1}{N_{s}} \sum_{i}^{N_s} p_i S_i(t) \right>\right],
\end{equation}
where angle brackets denote the statistical averaging, the square brackets are for
averaging over the different impurity configurations, and $N_s=pL^3$ is a number of spins in
the lattice. Two other interesting observables in short-time dynamics are the second moment
of magnetization $m^{(2)}(t)$,
\begin{equation} \label{m2}
m^{(2)}(t)=\left[ \left< \left( \frac{1}{N_{s}} \sum_{i}^{N_s} p_i S_i(t) \right)^2 \right>\right]
\end{equation}
and the auto-correlation function
\begin{equation} \label{A}
A(t)=\left[ \left< \frac{1}{N_{s}} \sum_{i}^{N_s} p_i S_i(t)S_i(0) \right>\right].
\end{equation}
As the spatial correlation length in the beginning of the time evolution is small, for a
finite system of dimension $d$ with lattice size $L$ the second moment
$m^{(2)}(t,L)\sim L^d$. Combining this with the result of the scaling form in Eq.~(\ref{mk})
for $\tau = 0$ and $b = t^{1/z}$, one obtains
\begin{eqnarray} \label{c2}
m^{(2)}(t) &\sim& t^{\, -2\beta/\nu z} m^{(2)}\left(1,t^{-1/z}L\right) \sim t^{\displaystyle \, c_2}, \\
       c_2 &=& \left(d-2\frac{\beta}{\nu}\right)\frac{1}{z}. \nonumber
\end{eqnarray}
Furthermore, careful scaling analysis shows that the auto-correlation also decays with a
power law \cite{Janssen92}
\begin{equation} \label{ca}
A(t) \sim t^{\displaystyle \, -c_a}, \qquad
 c_a = \frac{d}{z}-\theta'.
\end{equation}
Thus, the investigation of the short-time evolution of system from a high-temperature
initial state with $m_0=0$ allows to determine the dynamic exponent $z$, the ratio
of static exponents $\beta/\nu$, and the initial slip exponent $\theta'$.

Until now, a completely disordered initial state has been considered as starting
point, i.e., a state of very high temperature. The question arises how a completely
ordered initial state evolves, when heated up suddenly to the critical temperature.
In the scaling form (\ref{mk}), one can skip besides $L$, also the argument $m_0=1$,
\begin{equation}
m^{(k)}(t,\tau)=b^{-k\beta/\nu}m^{(k)}\left(b^{-z}t,b^{1/\nu}\tau \right).
\end{equation}
The system is simulated numerically by starting with a completely ordered state,
whose evaluation is measured at or near the critical temperature.
The quantities measured are $m(t)$ and $m^{(2)}(t)$. With $b=t^{1/z}$, one avoids the main
$t$ dependence in $m^{(k)}(t)$ and for $k=1$ one has
\begin{eqnarray} \label{moder}
m(t,\tau)&=&t^{-\beta/\nu z}m(1,t^{1/\nu z}\tau) \\
         &=& t^{-\beta/\nu z}\left(1+at^{1/\nu z}\tau+O(\tau^2)\right). \nonumber
\end{eqnarray}
For $\tau=0$, the magnetization decays by a power law $m(t)\sim t^{-\beta/\nu z}$.
If $\tau \neq 0$, the power law behavior is modified by the scaling function
$m(1,t^{1/\nu z} \tau)$. From this fact, the critical temperature $T_c$ and
the critical exponent $\beta/\nu z$ can be determined.

The scaling form of magnetization in Eq.~(\ref{moder}) is presented as follows:
\begin{equation} \label{logm}
\ln  m(t,\tau) = (-\beta/\nu z) \ln t + \ln m(1,t^{1/\nu z}\tau)
\end{equation}
after differentiation with respect to $\tau$ gives the power law of time dependence for the
logarithmic derivative of the magnetization in the following form:
\begin{equation} \label{logderm}
\left.\partial_{\tau} \ln  m(t,\tau) \right|_{\tau=0} \sim t^{1/\nu z},
\end{equation}
which allows to determine the ratio $1/\nu z$. On the basis of the magnetization and its
second moment, the cumulant
\begin{equation} \label{bcum}
U_2(t)= \frac{m^{(2)}}{(m)^2} - 1 \sim t^{d/z}
\end{equation}
is defined. From its slope, one can directly measure the dynamic exponent $z$.
Consequently, from an investigation of the system relaxation from ordered initial state
with $m_0=1$, the dynamic exponent $z$ and the static exponents $\beta$ and $\nu$
can be determined and their values can be compared to results of simulation of system
behavior from disordered initial state with $m_0=0$.

\section{Measurements of the critical exponents for 3D site-diluted Ising model}

We have performed simulations on three-dimensional cubic lattices with linear
size $L=128$, starting either from an ordered state or from a high-temperature
state with zero or small initial magnetization. We would like to mention
that measurements starting from a completely ordered state with the spins oriented
in the same direction ($m_0 = 1$) are more favorable, since they are much less affected
by fluctuations, because the quantities measured are rather big in contrast to those
from a random start with $m_0 = 0$. Therefore, for careful determination of the critical
exponents for 3D Ising model with spin concentrations $p=0.95$ and 0.8, we begin
to investigate the relaxation of this model from a completely ordered initial state.

\subsection{Evolution from an ordered state with $m_0 = 1$}

\begin{figure*}
\includegraphics[width=0.45\textwidth]{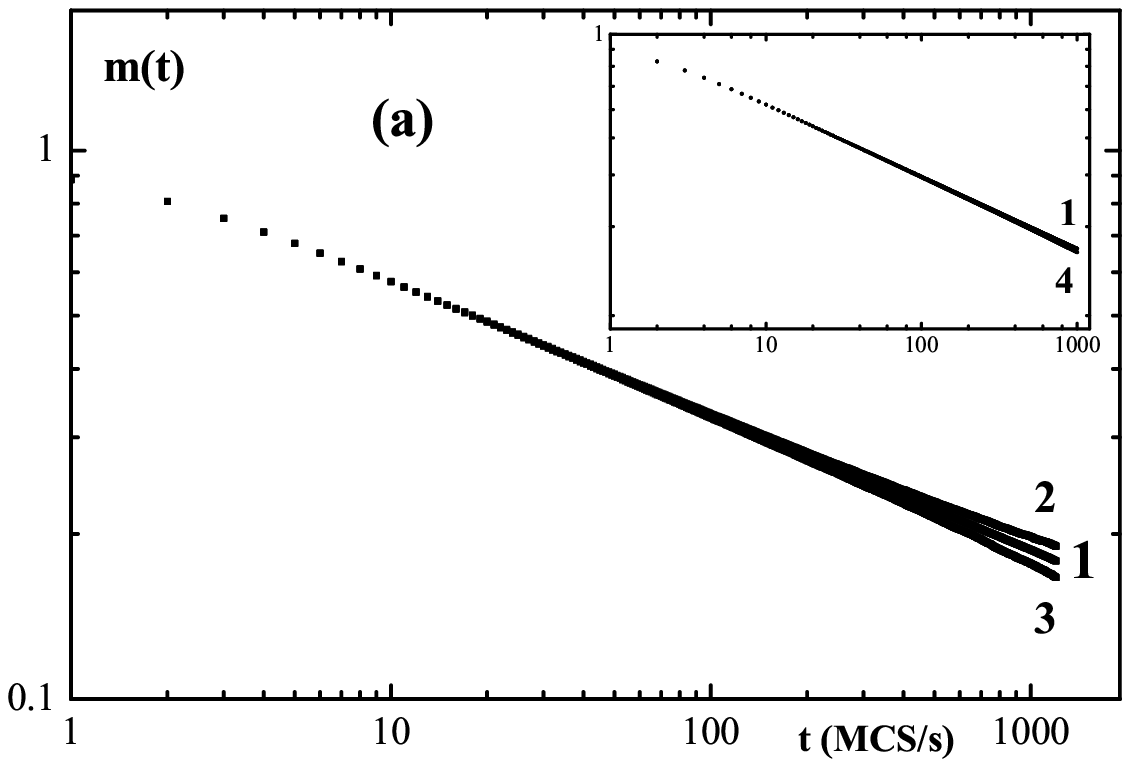}
\includegraphics[width=0.45\textwidth]{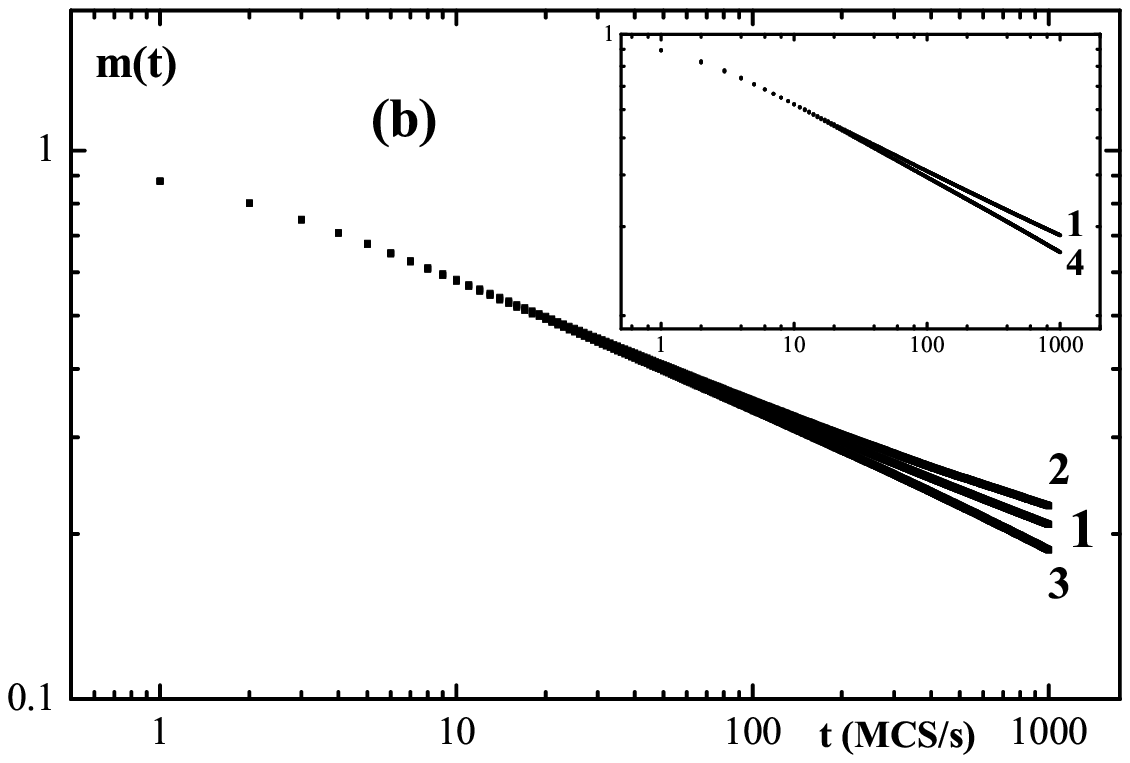}\\
\caption{\label{fig:2}
Time evolution of the magnetization $m(t)$ is plotted on a
log-log scale for samples with spin concentrations $p=0.95$ (a) and
$p=0.8$ (b) at $T=T_c(p)$ (curves 1) and at $T=T_c(p)\mp \Delta T$ with $\Delta T=0.005$
(curves 2 and 3). In insets curves 4 correspond to pure Ising model.
} \end{figure*}
\begin{figure*}
\includegraphics[width=0.45\textwidth]{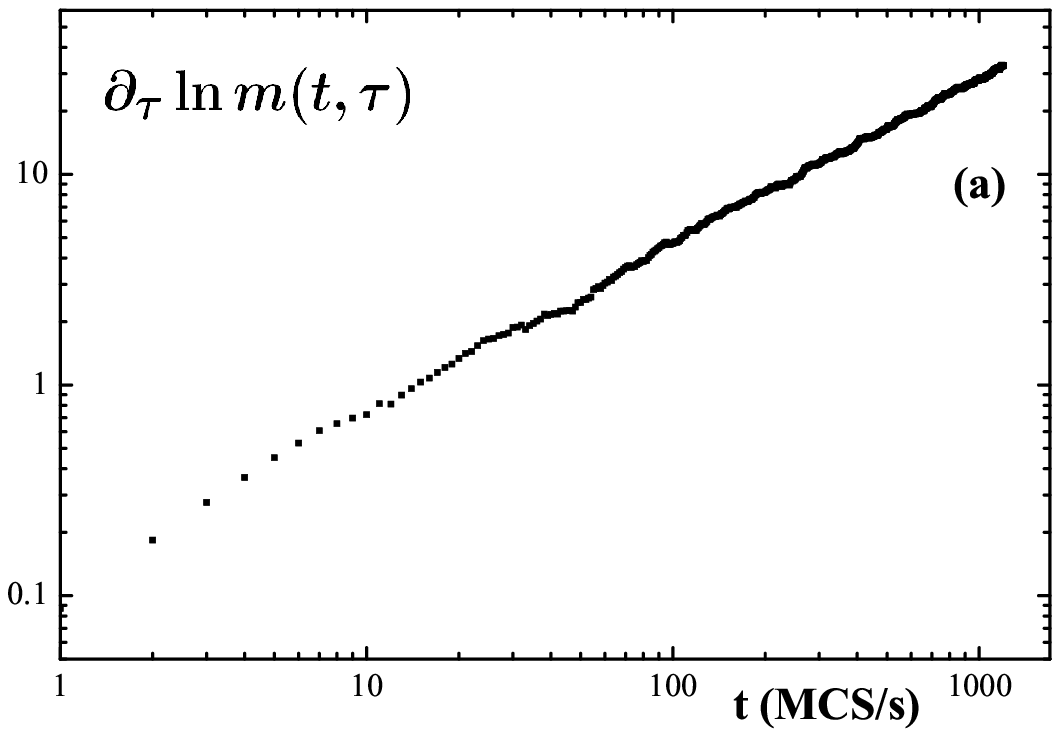}
\includegraphics[width=0.45\textwidth]{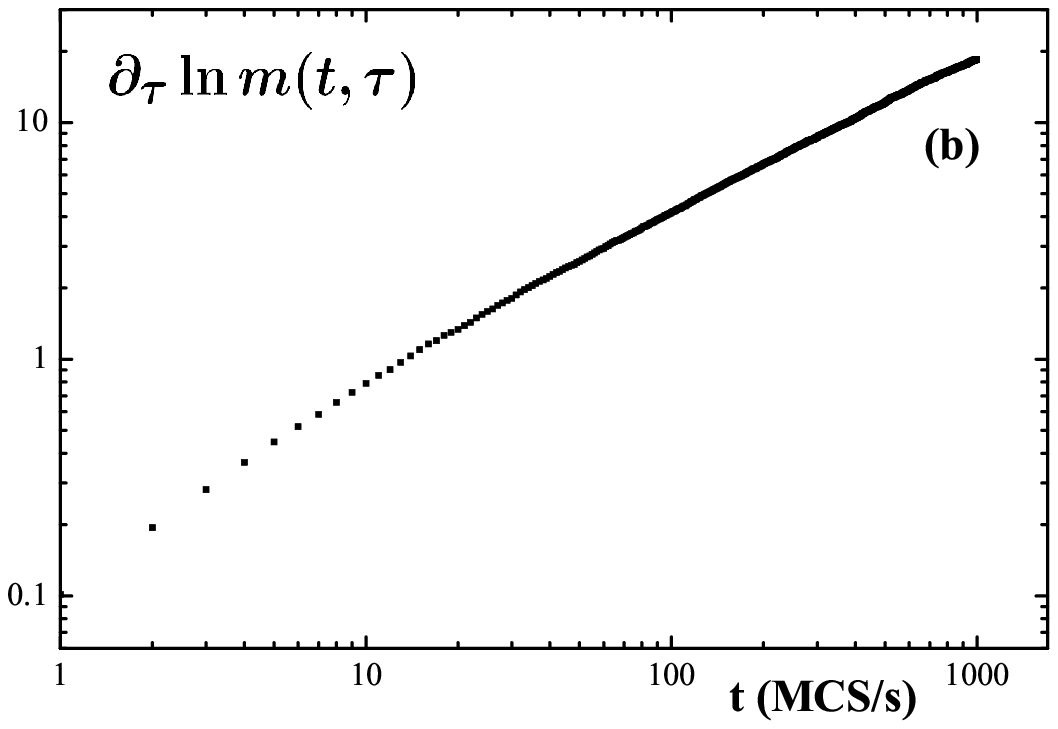}\\
\caption{\label{fig:3}
Time evolution of the logarithmic derivative of the magnetization $\left.\partial_{\tau} \ln  m(t,\tau) \right|_{\tau=0}$
with respect to $\tau$ is plotted on a log-log
scale for samples with spin concentrations $p=0.95$ (a) and $p=0.8$ (b).
} \end{figure*}
\begin{figure*}
\includegraphics[width=0.45\textwidth]{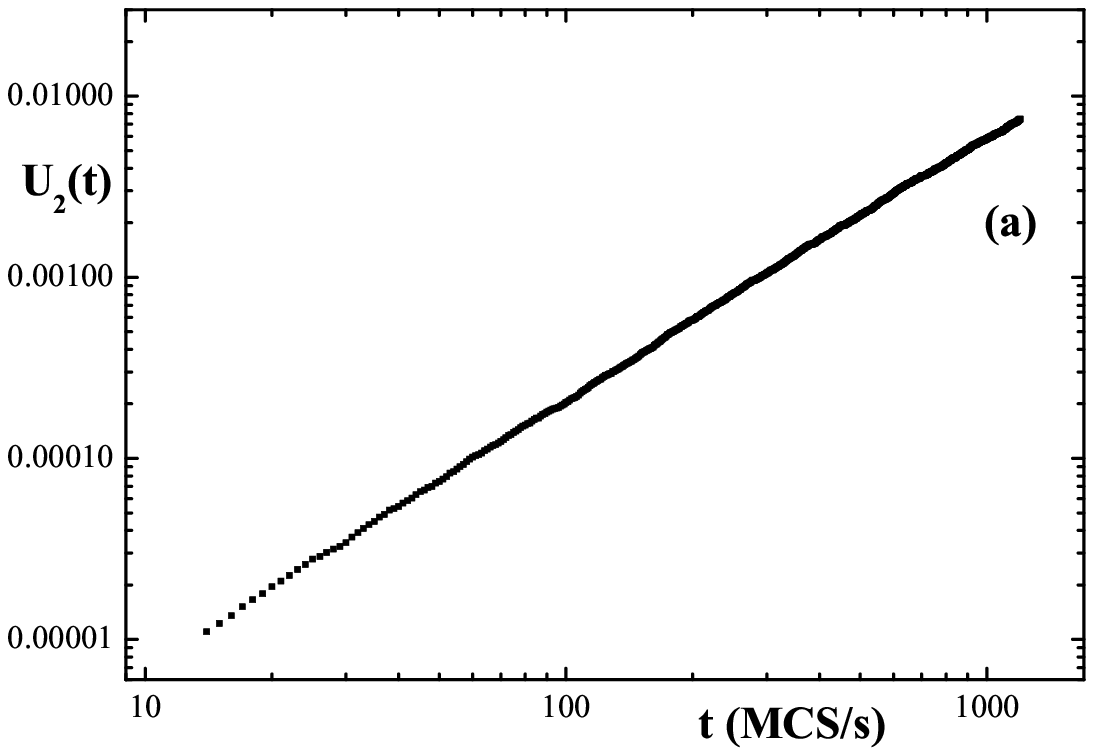}
\includegraphics[width=0.45\textwidth]{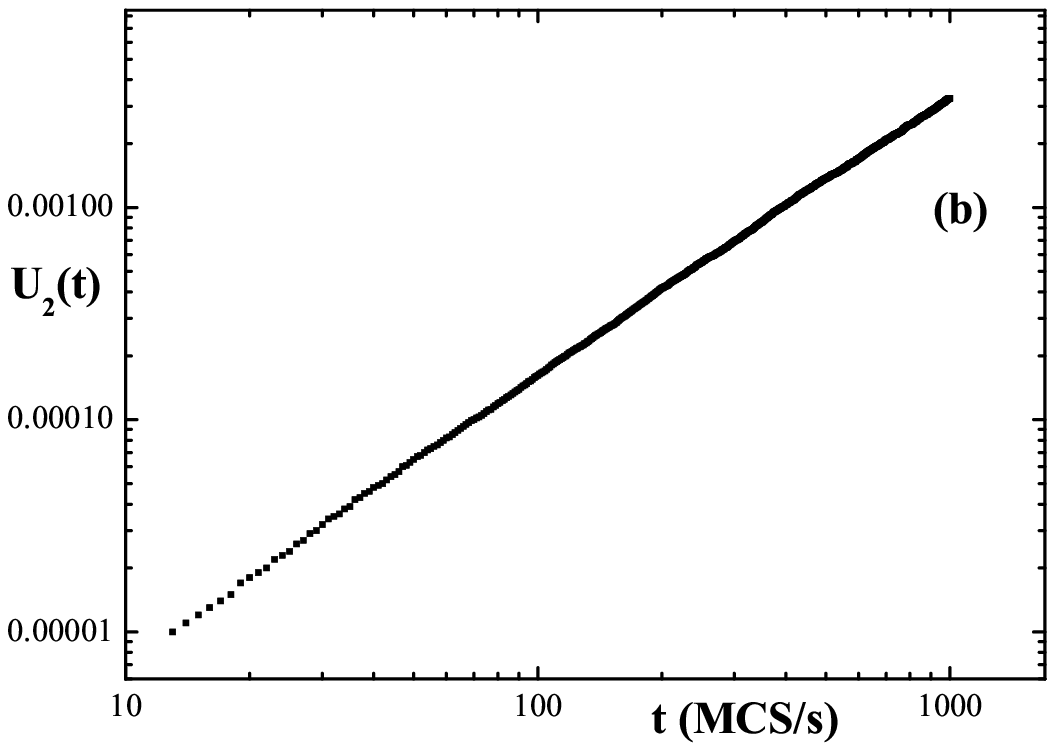}\\
\caption{\label{fig:4}
Time evolution of the cumulant $U_2(t)$ is plotted on a log-log
scale at $T=T_c(p)$ for samples with spin concentrations $p=0.95$ (a) and $p=0.8$ (b).
} \end{figure*}

\begin{figure*}
\includegraphics[width=0.31\textwidth]{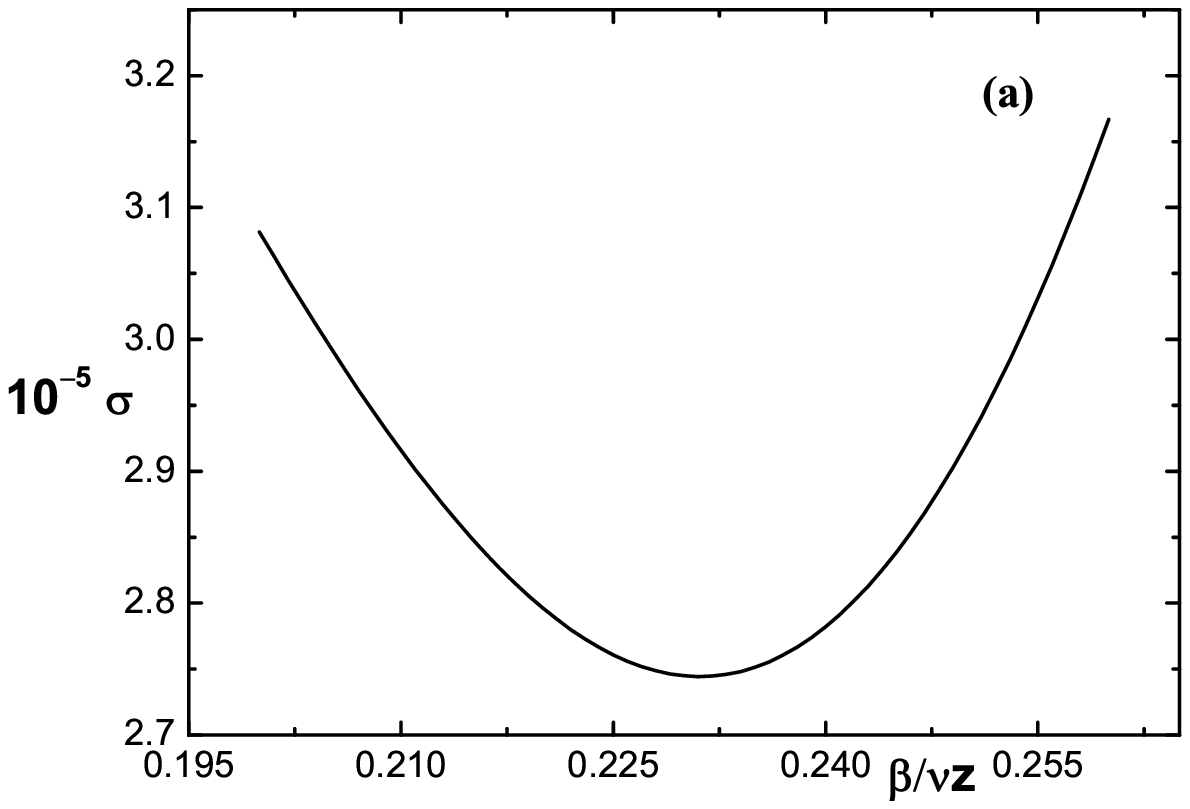}
\includegraphics[width=0.31\textwidth]{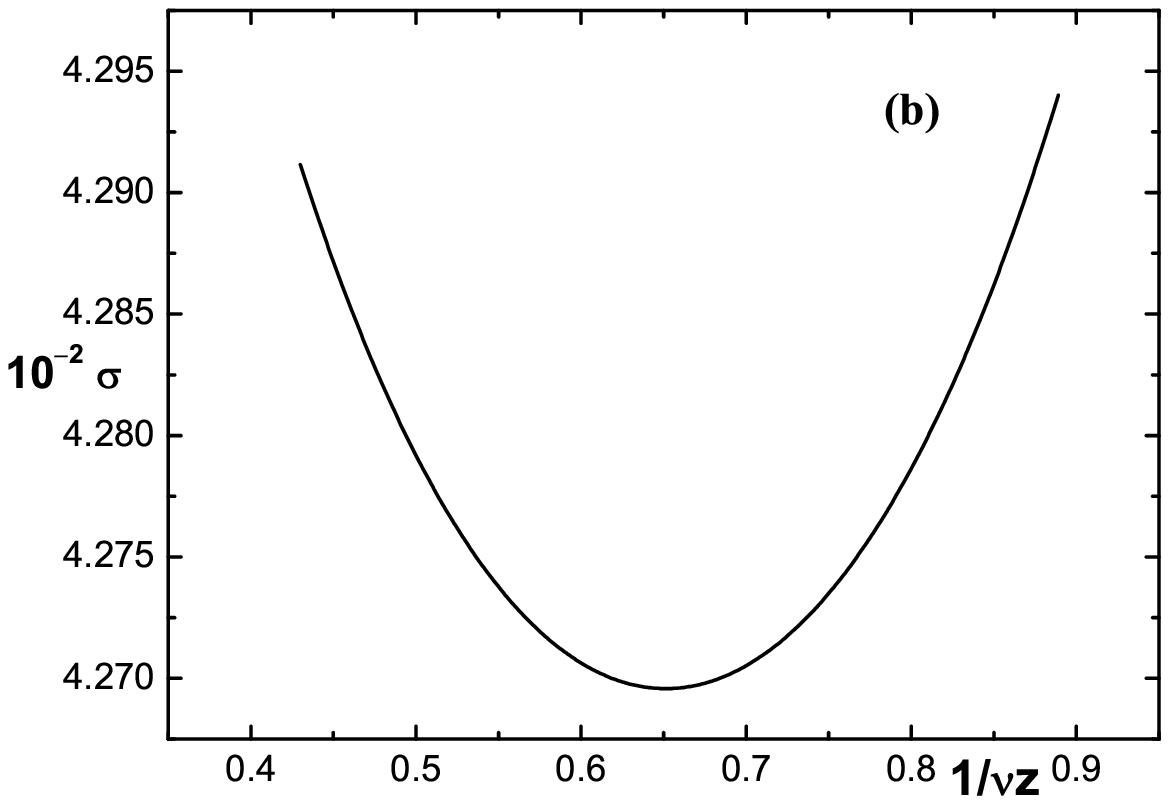}
\includegraphics[width=0.31\textwidth]{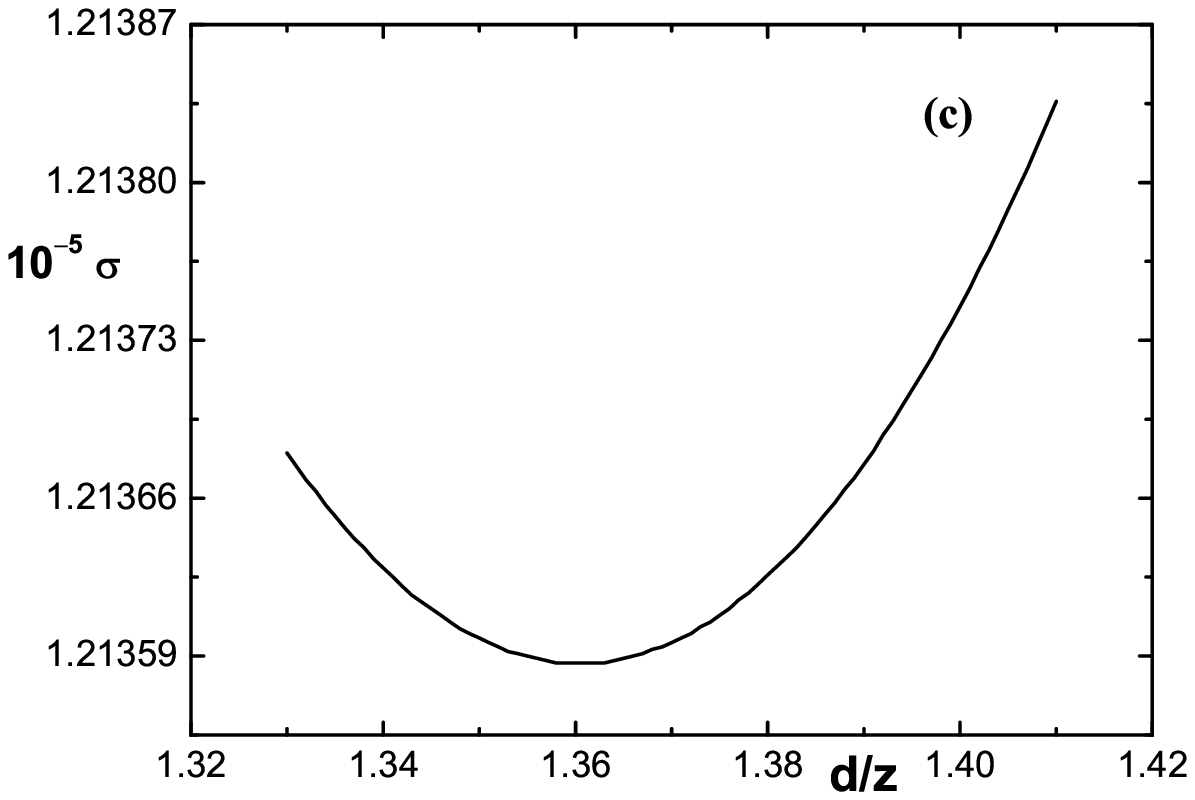}\\
\caption{\label{fig:5}
Dependence of the mean square errors
$\sigma$ of the fits for the magnetization (a), logarithmic
derivative of the magnetization (b),
and the cumulant (c) as a function of the
exponents $\beta/\nu z$, $1/\nu z$, and $d/z$ for $p = 0.8$.
} \end{figure*}

Initial configurations for systems with the spin concentrations $p=0.95; 0.8$ and  with
randomly distributed quenched pointlike defects were generated numerically. Starting from
those initial configurations, the system was updated with Metropolis algorithm at the
critical temperatures $T_c=4.26267(4)$ for $p=0.95$ and $T_c=3.49948(18)$ for $p=0.8$, which
have been determined in our paper \cite{Prudnikov2007} using Monte Carlo simulation of the 3D
site-diluted Ising model with different spin concentrations and particularly with $p = 0.95$
and 0.8 in equilibrium state. At present investigation, simulations have been performed up to
$t = 1000$ Monte Carlo steps per spin (MCS/s). We measured the time evolution of the magnetization $m(t)$ and the second
moment $m^{(2)}(t)$, which also allow to calculate the time-dependent cumulant
$U_2(t)$ in Eq.~(\ref{bcum}).

In Fig.~2 the magnetization $m(t)$ is plotted on a log-log scale for samples with
spin concentrations $p=0.95$ [Fig. 2, a] and $p=0.8$ [Fig. 2, b] at $T=T_c(p)$ (curves 1) and
at $T=T_c(p)\mp \triangle T$ with $\triangle T=0.005$ (curves 2 and 3). In Fig.~3,
the logarithmic derivative of the magnetization $\left.\partial_{\tau} \ln  m(t,\tau) \right|_{\tau=0}$
with respect to $\tau$ and in Fig.~4 the cumulant $U_2(t)$ are plotted on a log-log
scale at $T=T_c(p)$ for samples with spin concentrations $p=0.95$ (a) and $p=0.8$ (b),
accordingly. The $\left.\partial_{\tau} \ln  m(t,\tau) \right|_{\tau=0}$ have been obtained
from a quadratic interpolation between the three curves of time evolution of the magnetization
in Fig.~2 for the temperatures $T=T_c(p)$, $T=T_c(p)\mp \triangle T$ and taken at the critical
temperature $T_c=4.26267(4)$ for samples with $p=0.95$ and $T_c=3.49948(18)$ for $p=0.8$.
The resulting curves in Figs.~2-4 have been obtained by averaging over 6000 and 20000 samples
with different configurations of defects for systems with spin concentrations $p=0.95$ and
$p=0.8$, accordingly.

We have analyzed the time dependence of the cumulant $U_2(t)$ for samples with spin
concentrations $p=0.8$ and clarified that in the time interval $t \in [10,50]$ MCS/s
the $U_2(t)$ is best fitted by power law with the dynamic exponent $z=2.068(24)$, corresponding
to the pure Ising model \cite{Jaster,Prudnikov_JETPL97} and the influence of defects is
developed for $t > 400$ MCS/s only. An analysis of the $U_2(t)$ slope measured in the interval
$t\in [500,950]$ MCS/s shows that the exponent $d/z = 1.268(15)$ which gives $z = 2.366(28)$.
We have taken into account these dynamic crossover effects for the analysis of the time
dependence of magnetization and its derivative. So, the slope of magnetization measured in
the interval $t\in [400,950]$ MCS/s and its derivative over the interval $t\in [500,950]$ MCS/s
provides the exponents $\beta/\nu z=0.213(2)$ and $1/\nu z=0.600(8)$ which give
$\nu=0.704(18)$ and $\beta=0.365(8)$. The same analysis of the observable variables for samples
with spin concentrations $p=0.95$ leads to the value of exponent $d/z = 1.475(12)$, with
$z=2.034(16)$, in the time interval $t \in [10,200]$ MCS/s and to the exponents $d/z = 1.369(13)$,
$\beta/\nu z=0.213(2)$, and $1/\nu z=0.600(8)$ in the time interval $t \in [550,950]$ MCS/s,
which give $z=2.191(21)$, $\nu=0.704(18)$, and $\beta=0.365(8)$.

For demonstration of crossover effects between the pure and the dilute regimes,
we inserted in Fig.~2(a) and 2(b) results of the magnetization measurements for pure Ising model
at the critical temperature $T_c = 4.51142$ \cite{LandauFerrenberg}. Comparison of obtained
curves in the insets confirms our conclusion that the influence of disorder on the
non-equilibrium critical relaxation is developed for $t > 550$ MCS/s for samples
with $p=0.95$ and for $t > 400$ MCS/s for samples with $p=0.8$.

In the next stage, we have considered the corrections to the scaling in order
to obtain accurate values of the critical exponents. We have applied the following
expression for the observable $X(t)$:
\begin{eqnarray} \label{f_m}
X(t) = A_x t^{\delta}(1+B_x t^{-\omega/z}),
\end{eqnarray}
where $\omega$ is a well-known exponent of corrections to scaling, $A_x$ and $B_x$ are fitting
parameters, and an exponent $\delta=-\beta/\nu z$ when $X\equiv m(t)$, $\delta=d/z$ when
$X\equiv U_2(t)$, and $\delta=1/\nu z$ when $X\equiv \left.\partial_{\tau} \ln  m(t,\tau) \right|_{\tau=0}$.
This expression reflects the scaling transformation in the critical range of
time-dependent corrections to scaling in the form of $t^{-\omega/z}$ to the usual form of
corrections to scaling $\tau^{\omega \nu}$ in equilibrium state for time $t$ comparable to
the order parameter relaxation time $t_r \sim \xi^z \Omega(k\xi)$ \cite{H-H}. Field-theoretic
estimate of the $\omega$ value gives $\omega \simeq 0.25(10)$ in the six-loop approximation
\cite{Pelissetto}. Monte Carlo studies show that $\omega \simeq 0.370(63)$ from
Ref.~\cite{Ballesteros} and $\omega \simeq 0.26(13)$ from Ref.~\cite{Prudnikov2007}.

We have used the least-squares method for the best approximation of
the simulation data $X(t)$ by the expression in Eq.~(\ref{f_m}).
Minimum of the mean square errors $\sigma$ of this fitting procedure
determines the exponents $\delta$ and $\omega/z$. As example, for
samples with spin concentrations $p=0.8$, we plot in Fig.~5 the
$\sigma$ for the magnetization (a) as a function of the exponent
$\beta/\nu z$ for $\omega/z=0.275$, logarithmic derivative of the
magnetization (b) as a function of the exponent $1/\nu z$ for
$\omega/z=0.142$, and the cumulant (c) as a function of the exponent
$d/z$ for $\omega/z=0.132$. In Table I we present the computed
values of the exponents $\beta/\nu z$, $d/z$, $1/\nu z$, and
$\omega/z$, corresponding minimal values of the mean square errors
$\sigma$ in these fits. The statistical errors for exponents are
estimated by dividing all data into five data sets. On the base of
these values of exponents and average value of $\omega/z$, we
determine the final values of the critical exponents $z=2.185(25)$,
$\beta/\nu=0.533(13)$, $\nu=0.668(14)$, $\beta=0.356(6)$, and
$\omega=0.369(96)$ for $p=0.95$, and  $z=2.208(32)$,
$\beta/\nu=0.508(17)$, $\nu=0.685(21)$, $\beta=0.348(11)$, and
$\omega=0.404(110)$ for $p=0.8$.

\begin{table}
\caption{Values of the exponents $\beta/\nu z$, $d/z$, $1/\nu z$,
and $\omega/z$, corresponding minimal values of the mean-square errors for spin
concentrations $p=0.95$ and $p=0.80$. \label{tab:1}}
\begin{ruledtabular}
\begin{tabular}{clcccc}
$p$    & \multicolumn{1}{c}{Exponent} & Mean     & Approximation & Statistical & $\omega/z$ \\[-2pt]
       &                              & value    & errors        & errors      &            \\ \hline
$0.95$ & $\beta/\nu z$  & $0.244$   & $0.00011$ & $0.00131$ & $0.234$ \\
       & $d/z$          & $1.373$   & $0.00938$ & $0.00642$ & $0.092$ \\
       & $1/\nu z$      & $0.685$   & $0.00117$ & $0.00583$ & $0.181$ \\ \hline
$0.80$ & $\beta/\nu z$  & $0.230$   & $0.00081$ & $0.00393$ & $0.275$ \\
       & $d/z$          & $1.359$   & $0.01209$ & $0.00785$ & $0.132$ \\
       & $1/\nu z$      & $0.661$   & $0.00418$ & $0.00700$ & $0.142$ \\
\end{tabular}
\end{ruledtabular}
\end{table}

The comparison of the obtained values of critical exponents shows their belonging to the
same class of universal critical behavior of the diluted Ising model which can be
characterized by the averaged critical exponents $z=2.196(17)$, $\nu=0.677(11)$,
$\beta=0.352(5)$, and $\omega=0.387(60)$.

\subsection{Evolution from a disordered state with $m_0 \ll 1$}

In this part of the paper, we present the numerical investigations
of the short-time critical dynamics of the 3D site-diluted Ising
model on the lattice with linear size $L=128$, starting from a
disordered state with small initial magnetizations $m_0=0.01$, 0.02,
and 0.03 for samples with spin concentration $p=0.8$ only. For
independent determination of the dynamic critical exponent $z$ and
the ratio of static exponents $\beta/\nu$ we investigate also a time
dependence of the second moment of magnetization $m^{(2)}(t)$ and
the auto-correlation function $A(t)$ for system, starting from a
high-temperature initial state with $m_0=0$ (in fact, with
$m_0=10^{-4}$). In accordance with Sec.~II, a generalized dynamic
scaling predicts in this case a power law evolution for the
magnetization $m(t)$, the second moment $m^{(2)}(t)$ and the
autocorrelation function $A(t)$ in the short-dynamic regime.

Initial configurations for systems with the initial magnetization $m_0$ were generated
numerically. The initial magnetization has been prepared by flipping in an ordered state
a definite number of spins at randomly chosen sites in order to get the desired small value
of $m_0$. Starting from those initial configurations, the system was updated with Metropolis
algorithm at the critical temperature $T_c=3.49948(18)$, which has been determined in our
paper \cite{Prudnikov2007} using Monte Carlo simulation of the 3D site-diluted Ising model
with $p = 0.8$ in equilibrium state.

\begin{figure}
\includegraphics[width=0.45\textwidth]{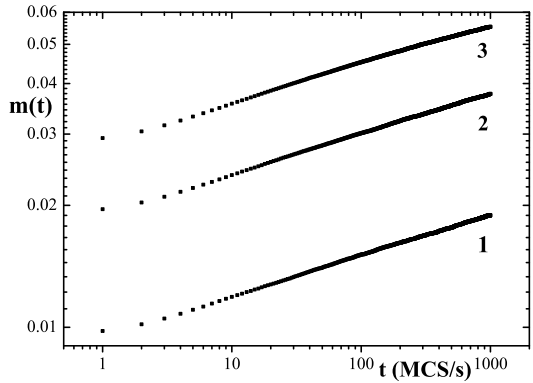}\\
\caption{\label{fig:6}
Time evolution of the magnetization $m(t)$ for different
values of the initial magnetization $m_0=0.01$ (1); $0.02$ (2); $0.03$ (3),
plotted on a log-log scale for samples with spin concentration $p=0.8$.\\[5mm]
} \end{figure}

\begin{figure*}
\includegraphics[width=0.45\textwidth]{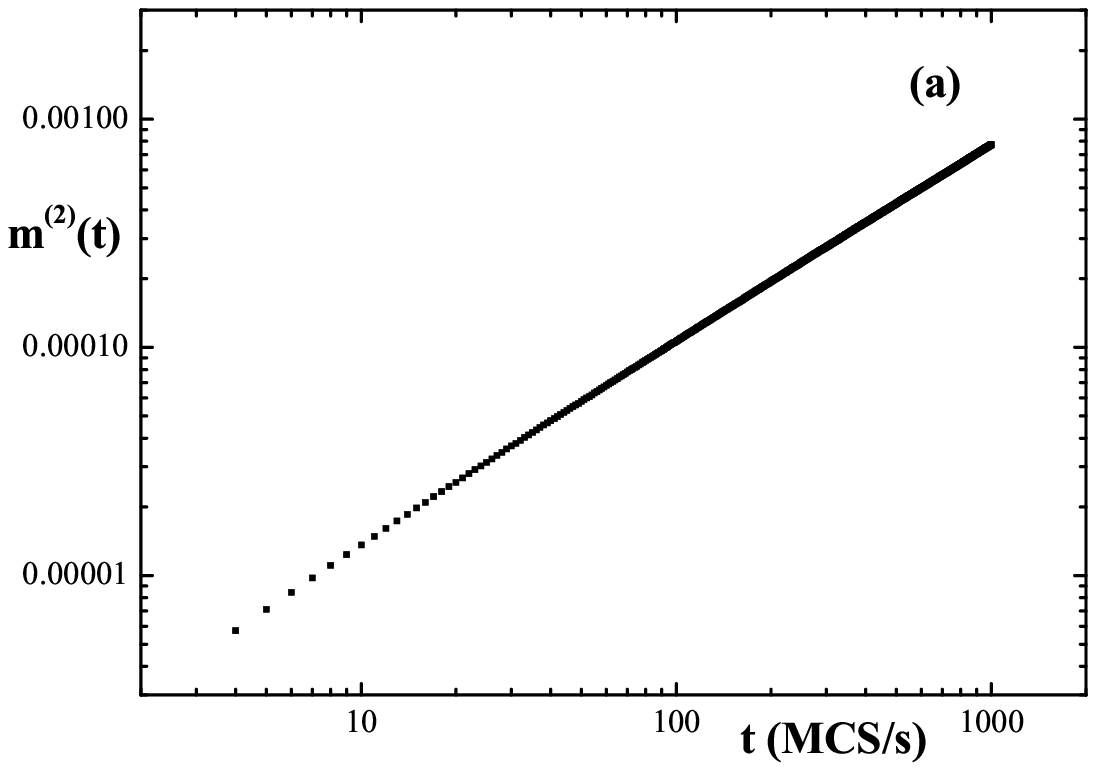}
\includegraphics[width=0.45\textwidth]{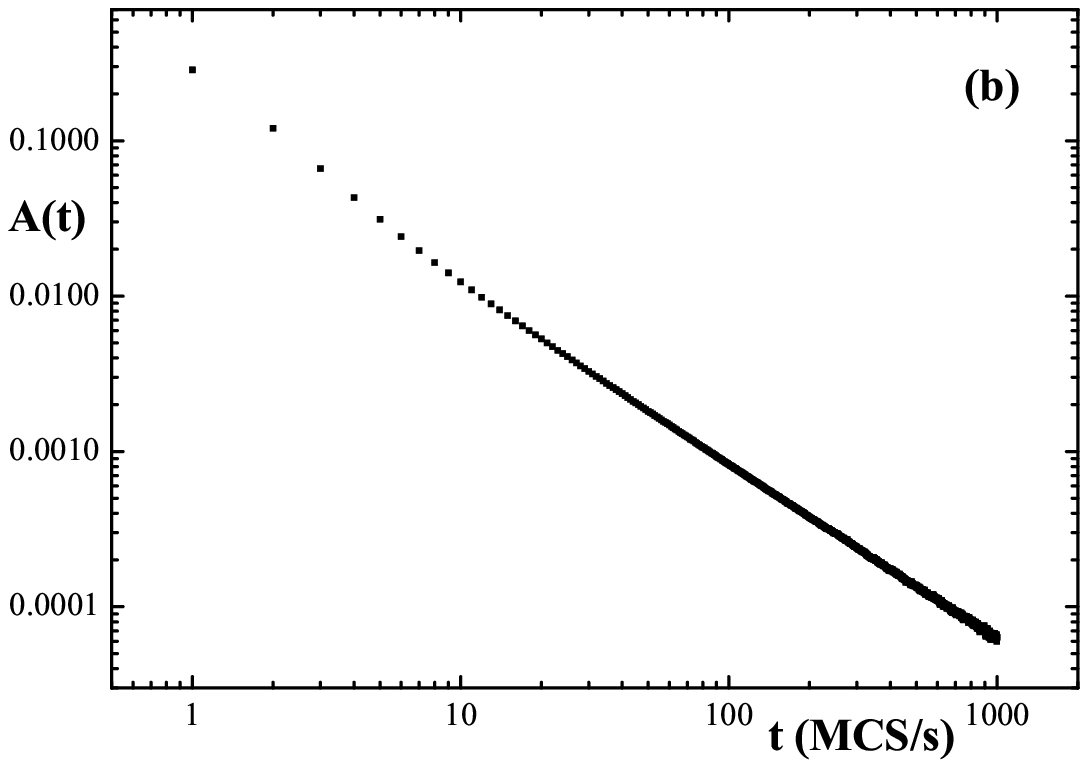}\\
\caption{\label{fig:7}
Time evolution of the second moment $m^{(2)}$
(a) and the correlation function $A(t)$ (b) for $L=128$ with the
initial magnetization $m_0=0.0001$. } \end{figure*}

\begin{table}
\caption{The initial slip exponent $\theta'$ measured by simulation of the
3D site-diluted Ising model with $p=0.80$ for different values of the initial
magnetization $m_0$ and exponents $c_2$ and $c_a$ for $m_0=0$.
The value $\theta'(m_0=0)$ is the result of an extrapolation. \label{tab:2}}
\begin{ruledtabular}
\begin{tabular}{c|ccccc}
  {$m_0$}  & {$\theta'$}    & {$c_2$}      & {$c_a$}       & $z$           & $\beta/\nu$     \\ \hline
           & {$t \in [15,60]$} & \multicolumn{2}{c}{$t \in [5,30]$}            \\ \hline
  {$0.03$} & {$0.1016(9)$}  &  &   \\
  {$0.02$} & {$0.1031(10)$} &  &   \\
  {$0.01$} & {$0.1043(12)$} &  &   \\
  {$0$}    & {$0.1057(17)$} & {$0.936(4)$} & {$1.347(8)$}  & {$2.065(14)$} & {$0.534(6)$} \\ \hline
           & {$t \in [300,800]$} & \multicolumn{2}{c}{$t \in [150,800]$}        \\ \hline
  {$0.03$} & {$0.083(3)$}   &  &   \\
  {$0.02$} & {$0.099(5)$}   &  &   \\
  {$0.01$} & {$0.105(9)$}   &  &   \\
  {$0$}    & {$0.122(11)$}  & {$0.859(5)$} & {$1.135(10)$} & {$2.387(20)$} & {$0.475(14)$} \\
\end{tabular}
\end{ruledtabular}
\end{table}

We measured the time evolution of the magnetization $m(t)$ with values of the initial
magnetization $m_0=0.01$, 0.02, and 0.03, the second moment $m^{(2)}(t)$ and the autocorrelation
function $A(t)$ with $m_0=0.0001$ up to $t = 1000$ MCS/s.
We show the obtained curves for $m(t)$ in Fig.~6, for $m^{(2)}(t)$
in Fig.~7a, and for $A(t)$ in Fig.~7b, which are plotted on a log-log scale. These curves were
obtained by averaging over 4000 different samples with 25 runs for each sample. We can see an
initial increase of the magnetization, which is a very prominent phenomenon in the short-time
critical dynamics. But in contrast to dynamics of the pure systems \cite{Jaster}, we can
observe the crossover from dynamics of the pure system on early times of the magnetization
evolution from $t\simeq 15$ up to $t\simeq 60$ MCS/s to dynamics of the disordered system with
the influence of point-like defects in the time interval $t\in[300,800]$ MCS/s. The same
crossover phenomena were observed in evolution of the second moment of magnetization
$m^{(2)}(t)$ and the autocorrelation function $A(t)$. In the result of linear approximation
of these curves in both the time intervals we obtained the values of the exponent
$\theta' (m_0)$ for initial states with $m_0 = 0.01$, $0.02$, and $0.03$ and the exponents $c_2$
and $c_a$ in accordance with relations in Eqs.~(\ref{theta'}), (\ref{c2}), and (\ref{ca})
(Table II). The final value of $\theta'$ is determined by extrapolation to $m_0 = 0$. Note
that the similar crossover phenomena in the non-equilibrium critical relaxation of systems
with quenched disorder have been revealed earlier in Ref.~\cite{PrudnikovPTP} by means of
numerical simulations of the critical behavior of the 3D Ising and XY models with linear
defects.

In the next stage, we have applied the procedure of corrections to the scaling determining by
the expression (\ref{f_m}) for analysis of the observable $m(t)$, $m^{(2)}(t)$, and $A(t)$.
We have used the least-squares method for the best approximation of the simulation
data by the expression (\ref{f_m}). Minimum of the mean-square errors $\sigma$ of this fitting
procedure determines the exponents $\theta'(m_0)$, $c_2$, and $c_a$ with their respective
$\omega/z$. In Table III, we present the computed values of these exponents and the final value
of $\theta'= 0.127(16)$ obtained by extrapolation of $\theta'(m_0)$ for different values of
the initial magnetization $m_0$ to $m_0=0$. In Table III, we also give the values of critical
exponents $z$, $\beta/\nu$, and the average value of $\omega$ and compare the values of these
exponents to values of corresponding exponents for the pure Ising model \cite{Jaster}.
The obtained values agree quite well with results of simulation from an ordered state with
$m_0=1$.

\begin{table}
\caption{Values of the exponents for the 3D site-diluted Ising model
to $p=0.80$ obtained with the use of corrections to the scaling. \label{tab:3}}
\begin{ruledtabular}
\begin{tabular}{ccc}
  Exponent             & Value & {$\omega/z$}  \\ \hline
  {$\theta'(m_0=0.03)$} & {$0.104(12)$} & {$0.074$} \\
  {$\theta'(m_0=0.02)$} & {$0.117(10)$} & {$0.068$} \\
  {$\theta'(m_0=0.01)$} & {$0.118(10)$} & {$0.096$} \\ \hline
  {$\theta'(m_0\to 0)$} & {$0.127(16)$} & {$0.079$}  \\
  {$c_2(m_0=0)$}        & {$0.909(4)$}  & {$0.112$}  \\
  {$c_a(m_0=0)$}        & {$1.242(10)$} & {$0.160$}  \\ \hline
  {$z$}                & {$2.191(21)$} \\
  {$\beta/\nu$}        & {$0.504(14)$} \\
  {$(\omega/z)_{\text{av}}$}                & {$0.117(24)$} \\
  {$(\omega)_{\text{av}}$}          & {$0.256(55)$} \\
\end{tabular}
\end{ruledtabular}
\end{table}

Comparison of the value $\theta'=0.127(16)$ to $\theta'= 0.10(2)$
from Ref.~\cite{Schehr} also measured by simulation of the 3D
site-diluted Ising model shows their not bad agreement within the
limits of statistical errors of simulation and numerical
approximations. In Ref.~\cite{Schehr} the non-equilibrium relaxation
of the magnetization $m(t)$ has been investigated from the initial
random spin configurations with mean magnetization $m_0=0.01$ only
for samples with different spin concentrations $p=0.499$, 0.6, 0.65,
and 0.8 and with linear sizes $L=8$, 16, 32 and 64. However, in
accordance with Ref.~\cite{Jaster}, the initial slip exponent
$\theta'$ must be determined in the asymptotical limit with $m_0 \to
0$ on the basis of results of $m(t)$ computing for a few small
values of the initial magnetization $m_0$. Furthermore, as it
follows from Eq.~(5), the magnetization undergoes a power­law
initial increase characterized by $\theta'$ for sufficiently small
$t^{x_0/z}m_0$. For $m_0$ and $t$ not too small, the power­law
behavior will be modified. For strongly diluted systems which are
also considered in Ref.~\cite{Schehr}, the influence of disorder is
observed for longer times than for weakly diluted systems.
Therefore, the use of the same value of the initial magnetization
for determination of the initial slip exponent $\theta'$ for both
weakly and strongly diluted systems is unjustified. It should be
noted that in Ref.~\cite{Schehr} the data analysis of $m(t)$ for
samples with different spin concentrations $p$ was carried out with
the use of the corrections to scaling procedure during realization
of which the universal value of the dynamic critical exponent
$z=2.62(7)$ obtained in Ref.~\cite{Parisi} was applied. But this
value $z$ is inconsistent with values calculated both in the present
paper and in Ref.~\cite{Prudnikov2006} in the three-loop
approximation of the field-theoretic RG description, and with
experimentally measured value $z=2.18(10)$ for weakly diluted Ising
magnet $\mathrm{Fe}_{0.9}\mathrm{Zn}_{0.1}\mathrm{F}_{2}$ from
Ref.~\cite{Rosov}.

Obtained in present paper is the asymptotic value $\theta'(m_0\to
0)=0.127(16)$ that demonstrates that it is larger than
$\theta'(m_0=0.01)$ from Ref.~\cite{Schehr}. It is explained by the
revealed tendency that $\theta'(m_0^{(2)}) > \theta'(m_0^{(1)})$ for
the initial magnetization which are in the following correspondence
with each other as $m_0^{(2)} < m_0^{(1)}$. Therefore, the
distinguished good agreement in Ref.~\cite{Schehr} of the obtained
value $\theta'= 0.10(2)$ with $\theta'\simeq 0.0867$ from
Ref.~\cite{Oerding} is unjustified. The results of investigations
carried out in this paper give reasons to consider that the value of
the initial slip exponent $\theta'=0.127(16)$ is more realistic for
description of non-equilibrium critical relaxation of the 3D weakly
diluted Ising-like systems which is larger than the value of
exponent $\theta'= 0.108(2)$ for the pure 3D Ising systems
\cite{Jaster,Prudnikov2008} rather than smaller as predicted by
results from Refs.~\cite{Schehr} and \cite{Oerding}.

We have realized a field-theoretic renormalization-group description of non-equilibrium critical
relaxation for directly three-dimensional diluted Ising model and calculated the initial slip
exponent $\theta'$ in two-loop approximation without using the $\varepsilon$-expansion method.
As a result, we have obtained
\begin{align}\label{th}
\theta' =\ &\frac{1}{\,6\,}\,g^*+0.125v^*-0.123968(g^*)^2+ \\ &+0.14680608g^*v^*-0.0156245(v^*)^2, \nonumber
\end{align}
where $g^*$ and $v^*$ are values of the vertexes describing
interaction of the order parameter fluctuations in the fixed-point
of the renormalization-group equations \cite{Amit,Zinn-Justin}. For
further calculations, we use the FP with $g^*=2.2514(42)$, $v^*=-
0.7049(13)$ which determines the critical behavior of the 3D dilute
Ising model. The coordinate of this FP has been obtained in our
paper \cite{Prudnikov2006} as average of numerical values $g^*$,
$v^*$ which were calculated with the use of different methods of
resummation technique. It is well known that the series expansions
for the critical exponents exhibit factorial divergence, but they
can be considered in an asymptotic context. In order to obtain
physically reasonable values of the critical exponents for 3D
systems, special methods for the summation of asymptotic series have
been developed
\cite{Baker,LeGuillou,Antonenko,Kazakov,Suslov,Prudnikov2006}, the
most effective being the Pad\'{e}-Borel, Pad\'{e}-Borel-Leroy (PBL), and
conformal mapping techniques. We employ the PBL resummation method
extended to the two-parameter case. The PBL
method is a generalization of the Pad\'{e}-Borel method with the
integral Borel transformation
\begin{equation}
\begin{split}
f(g)=\sum\limits_{n=0}^{\infty} c_n g^n =\int\limits_{0}^{\infty}dte^{-t}B(gt^{b}), \\ B(g)=\sum\limits_{n=0}^{\infty}B_{n}g^{n},\quad B_{n}=\frac{c_{n}}{\Gamma(bn+1)},
\label{f_f_beta}
\end{split}
\end{equation}
where the value of parameter $b=2.221426$ was chosen in Ref.~\cite{Prudnikov2006} from
convergence analysis of the test series for exactly solvable problem of calculating
the anharmonic oscillator energy with the asymptotic convergence of the series, which is
similar to the series for the RG $\beta$ and $\gamma$ functions in the theory of critical
phenomena. As a result, the following value $\theta'=0.1203$ has been calculated which
very well agrees with our results of simulation.

\section{Analysis of results and conclusions}

\begin{table*}
\caption{Values of the obtained critical exponents and comparison to other results of Monte Carlo simulations (MC),
 field-theoretical method with fixed-dimension $d=3$ expansion (FTM), and experimental (EXP) investigations}
\begin{ruledtabular}
\begin{tabular}{lc|llllll}
                                                                                 && \multicolumn{1}{c}{$z$}  & \multicolumn{1}{c}{$\theta'$} & \multicolumn{1}{c}{$\beta/\nu$} & \multicolumn{1}{c}{$\nu$} & \multicolumn{1}{c}{$\beta$} & \multicolumn{1}{c}{$\omega$}  \\ \hline
 $p=0.95$, $m_0=1$                                                               && $2.185(25)$ &             & $0.533(13)$ & $0.668(14)$ & $0.356(6)$  & $0.369(96)$   \\
 $p=0.80$, $m_0=1$                                                               && $2.208(32)$ &             & $0.508(17)$ & $0.685(21)$ & $0.348(11)$ & $0.404(110)$  \\
 $p=0.80$, $m_0\ll1$                                                             && $2.191(21)$ & $0.127(16)$ & $0.504(14)$ &             &             & $0.256(55)$   \\  \hline
 Pelissetto, Vicari, 2000,        (Ref.~\cite{Pelissetto})               &(FTM)&             &             & $0.515(15)$ & $0.678(10)$ & $0.349(5)$  & $0.25(10)$    \\
 Prudnikov, \emph{et al.}, 2006,       (Ref.~\cite{Prudnikov2006})              &(FTM)&  $2.1792(13)$ \\ \hline
 Rosov, \emph{et al.}, 1988,  $\mathrm{Fe}_{p}\mathrm{Zn}_{1-p}\mathrm{F}_{\!2}$ $p=0.9$, (Ref.~\cite{Rosov1988})       &(EXP)&  &  &  &  & $0.350(9)$   \\
 Rosov, \emph{et al.}, 1992,  $\mathrm{Fe}_{p}\mathrm{Zn}_{1-p}\mathrm{F}_{\!2}$ $p=0.9$, (Ref.~\cite{Rosov})           &(EXP)&  $2.18(10)$   \\
 Slani\u{c}, \emph{et al.}, 1999, $\mathrm{Fe}_{p}\mathrm{Zn}_{1-p}\mathrm{F}_{\!2}$ $p=0.93$, (Ref.~\cite{Slanic1999}) &(EXP)&             &             &             & $0.70(2)$ &             &               \\\hline
 Prudnikov, Vakilov, 1992, $p=0.95$,                                             &             &  $2.19(7) $ &             &             &             &             &               \\
 \hspace*{36mm}           $p=0.80$,                                              &             &  $2.20(8) $ &             &             &             &             &               \\
 \hspace*{36mm}           $p=0.60$,                                              &             &  $2.58(9) $ &             &             &             &             &               \\
 \hspace*{36mm}           $p=0.40$, (Ref.~\cite{Prudnikov92})              &(MC) &  $2.65(12)$ &             &             &             &             &               \\
 Heuer, 1993, $p=0.95$                                                           &             &  $2.16(1)$  &             & $0.49(2)$   & $0.64(2)$   & $0.31(2)$   &               \\
 \hspace*{18mm} $p=0.90$                                                         &             &  $2.232(4)$ &             & $0.48(2)$   & $0.65(2)$   & $0.31(2)$   &               \\
 \hspace*{18mm} $p=0.80$,                                                        &             &  $2.38(1)$  &             & $0.51(2)$   & $0.68(2)$   & $0.35(2)$   &               \\
 \hspace*{18mm} $p=0.60$, (Refs.~\cite{Heuer90},\cite{Heuer93})            &(MC) &  $2.93(3)$  &             & $0.45(2)$   & $0.72(2)$   & $0.33(2)$   &               \\
 Wiseman, Domany, 1998, $p=0.80$,                                                &             &             &             & $0.505(2)$  & $0.682(2)$  &             &               \\
 \hspace*{31.5mm} $p=0.60$, (Ref.~\cite{Wiseman})                          &(MC) &             &             & $0.437(21)$ & $0.717(6)$  &             &               \\
 Ballesteros, \emph{et al.}, 1998, $p=0.90\div 0.40$ (Ref.~\cite{Ballesteros})    &(MC) &             &             & $0.519(8)$  & $0.684(5)$  & $0.355(3)$  & $0.370(63)$   \\
 Parisi, \emph{et al.}, 1999, $p=0.90\div 0.40$ (Ref.~\cite{Parisi})              &(MC) & $2.62(7)$   &             &             &             &             & $0.50(13)$    \\
 Calabrese, \emph{et al.}, 2003, $p=0.80$ (Ref.~\cite{Calabrese})                 &(MC) &             &             & $0.518(5)$  & $0.683(3)$  & $0.354(2)$  &              \\
 Murtazaev, \emph{et al.}, 2004, $p=0.95$,                                              &             &             &             &             & $0.646(2)$  & $0.306(3)$  &               \\
 \hspace*{33mm} $p=0.9$,                                                         &             &             &             &             & $0.664(3)$  & $0.308(3)$  &               \\
 \hspace*{33mm} $p=0.8$,                                                         &             &             &             &             & $0.683(4)$  & $0.310(3)$  &               \\
 \hspace*{33mm} $p=0.6$, (Ref.~\cite{Murtazaev2004})                       &(MC) &             &             &             & $0.725(6)$  & $0.349(4)$  &               \\
 Schehr, Paul, 2006, (Ref.~\cite{Schehr})                                  &(MC) &             & $0.10(2)$ \\
 Hasenbusch, \emph{et al.}, 2007, $p=0.8$ (Ref.~\cite{Hasenbusch})                &(MC) & $2.35(2)$   &  \\
 Prudnikov, \emph{et al.}, 2007, $p=0.95\div0.80$,                                      &             &             &             & $0.532(12)$ & $0.693(5)$  &             & $0.26(13)$    \\
 \hspace*{33mm}           $p=0.60\div0.50$, (Ref.~\cite{Prudnikov2007})    &(MC) &             &             & $0.524(13)$ & $0.731(11)$ &             & $0.28(15)$    \\
\end{tabular}
\end{ruledtabular}
\end{table*}

In a summary Table IV, we present the values of critical exponents $z$, $\theta'$,
$\beta/\nu$, $\nu$, $\beta$, and $\omega$ obtained in this paper by comprehensive Monte
Carlo simulations of the short-time critical evolution of the 3D site-diluted Ising model
both from an ordered initial state with $m_0=1$ for samples with spin concentrations $p=0.95$
and 0.8 and from a disordered initial states with $m_0 \ll 1$ for samples with spin
concentration $p=0.8$. For comparison, we give in Table IV the results of calculation of these
exponents by the field-theoretical method with fixed-dimension $d=3$ expansion
\cite{Pelissetto,Prudnikov2006}, the results of experimental investigations of the Ising-like
magnets \cite{Rosov,Slanic1999,Rosov1988}, and the results of numerical studies
\cite{Ballesteros,Prudnikov2007,Schehr,Parisi,Wiseman,Calabrese,Heuer90,Heuer93,Prudnikov92,Murtazaev2004,Hasenbusch}.
As shown in Table IV, our values of exponents are in good agreement within the limits of
statistical errors of simulation and numerical approximations with results of the
field-theoretical description of the statics in six-loop approximation \cite{Pelissetto} and
critical dynamics in three-loop approximation \cite{Prudnikov2006} and with the results of
experimental investigations of the static \cite{Slanic1999,Rosov1988} and dynamic \cite{Rosov} critical
behaviors of weakly diluted Ising-like magnets.

Comparison to results of Monte Carlo simulations shows that our static exponents $\beta/\nu$,
$\nu$, and $\beta$ agree well with those exponents measured in equilibrium for weakly diluted
systems in the most cited papers and for systems with wide dilution range in
Ref.~\cite{Ballesteros} where the critical exponents were obtained for $p\leq 0.8$ as
dilution-independent after a proper infinite volume extrapolation with taking into account
the leading corrections-to-scaling terms. However, it was found that the case with $p=0.9$
falls out from this dilution-independent scheme of fits with common exponent $\omega=0.37$
for samples with different spin concentrations. Authors draw a conclusion that the $p=0.9$
data seem to be still crossing over from the pure Ising fixed point to the diluted one.
Most of the computations have been carried out at $p=0.8$ as in this case the scaling
corrections are very small and the results even in small lattices are stable. So, in
Ref.~\cite{Calabrese}, it was shown that for case with $p=0.8$, the observed corrections
to scaling could be the next-to-leading with $\omega_2 \simeq 0.80$. However, our present
investigation and the results of computations in equilibrium \cite{Prudnikov2007} show that
the systems with $p=0.95$ and $p=0.8$ are characterized by close agreement of the critical
exponent values and they belong to the same class of universal critical behavior
of the site-diluted Ising model with the averaged critical exponents
$\nu=0.677(11)$, $\beta=0.352(5)$, and $\omega=0.387(60)$.

Now we compare the values of the dynamic critical exponent $z$ obtained in this paper by
the short-time dynamics method to the results of Monte Carlo simulations of the critical
dynamics in equilibrium, realized in Refs.~\cite{Heuer93} and \cite{Hasenbusch}, to the
results of non-equilibrium studies of the susceptibility in Refs.~\cite{Parisi} and \cite{Hasenbusch},
and to the results of Monte Carlo renormalization group application to description of the
site-diluted Ising model relaxation from the ordered initial state with $m_0=1$ in
Ref.~\cite{Prudnikov92}. Values of $z$ from paper \cite{Heuer93} agree rather well
with our results only for weakly diluted systems with $p \geq 0.9$, while a noticeable
difference between the results is observed for strongly disordered systems.
Starting from the universality concept for critical behavior of diluted Ising systems
and that the asymptotic value of $z$ is independent of the degree of
dilution, the author in Ref.~\cite{Heuer93} obtained the asymptotic value $z=2.4(1)$
using the effective values of the exponent listed in Table IV.
The off-equilibrium critical dynamics of the 3D Ising model with the spin concentration
varying in a wide range was analyzed in Ref.~\cite{Parisi}. Also assuming that
the critical behavior of diluted Ising systems is universal under dilution, the authors
obtained the asymptotic value of $z=2.62(7)$ taking into account the leading corrections
to the scaling dependence for the dynamical susceptibility. In this case, the value of the
exponent $\omega = 0.50(13)$ obtained in Ref.~\cite{Parisi} is strongly inconsistent
with $\omega = 0.25(10)$ from the field theory calculations \cite{Pelissetto} and not so well
agreement with $\omega = 0.37(6)$ from Monte Carlo results in Ref.~\cite{Ballesteros}.
In the approximations realized in Ref.~\cite{Parisi}, the results for weakly diluted
systems were characterized by the largest errors.
In Ref.~\cite{Hasenbusch}, it was carried out the Metropolis dynamics in equilibrium
for site-diluted Ising model with $p=0.8$, 0.85, and 0.65.
For case with $p=0.8$, authors investigated in detail the scaling corrections which
were the next-to-leading with $\omega_2 = 0.82(8)$ and gave the exponent $z=2.35(2)$.
Investigations for other values of $p$ did not permit to determine $z$ in these systems
accurately. Also, in Ref.~\cite{Hasenbusch}, investigated was    the off-equilibrium
relaxational critical dynamics in the site-diluted Ising model at $p = 0.8$.
The results show that equilibrium estimate $z = 2.35(2)$ is perfectly consistent with
the off-equilibrium MC data. Authors did not observe a large-time scaling corrections
proportional to $t^{-\omega/z}$; instead, their data show corrections that are proportional
to $t^{-\omega_2/z}$ with the static correction-to-scaling exponents $\omega = 0.29(2)$
and $\omega_2 = 0.82(8)$.
We have some doubts in validity of $z=2.35(2)$. As it was shown in \cite{Prudnikov2007},
the realization of correction to scaling procedure demands the six simulation data points
at least for their approximation by four-parameter function such as in Eq.~(\ref{f_m}) for lattices with
$L>L_{\mathrm{min}}$. Whereas in paper \cite{Hasenbusch}, the asymptotical value $z=2.35(2)$ was
obtained with the use of four or five data points as it was demonstrated in Figs.~1-4 in
\cite{Hasenbusch}. It is necessary to use additional data points for lattices with $L>64$.

The early results of our numerical investigations of the critical dynamics for diluted Ising
systems in Ref.~\cite{Prudnikov92} by Monte Carlo renormalization-group method show
very good agreement with our present results for weakly diluted systems, while a noticeable
difference between the results is observed for strongly diluted systems.
We are planning to continue the Monte Carlo study of critical behavior
of the site-diluted Ising model by short-time dynamics method with $p=0.6$ and $0.5$
focusing on the problem of dilution independence of asymptotic characteristics.

The present results of Monte Carlo investigations allow us to recognize that the
short-time dynamics method is reliable for the study of the critical behavior of the
systems with quenched disorder and is the alternative to traditional Monte Carlo
methods. But in contrast to studies of the critical behavior of the pure systems by
the short-time dynamics method \cite{Jaster,Zheng}, in case of the systems with quenched
point-like disorder after the microscopic time $t_{mic} \simeq 5\div10$ MCS/s,
there exist three stages of dynamic evolution. For systems starting from the ordered
initial states ($m_0 = 1$) in the time interval of 10-50 MCS/s, the power-law dependence
is observed at the critical point for Binder cumulant $U_2(t)$, which is similar to that
in the pure system. In the time interval [400,950], the power-law dependences are observed
in the critical point for the magnetization $m(t)$, the logarithmic derivative of the
magnetization, and cumulant $U_2(t)$ which are determined by the influence
of disorder. However, careful analysis of the slopes for $m(t)$ and
$U_2(t)$ reveals that a correction to scaling should be considered in order to obtain
accurate results. The dynamic and static critical exponents were computed with the
use of the corrections to scaling, which demonstrate their good agreement with results
of the field-theoretic description of the critical behavior of these models with disorder.
In the intermediate time interval of 100–-400 MCS/s, the dynamic crossover behavior is observed
from the critical behavior typical for the pure systems to behavior determined
by the influence of disorder.

The investigation of the critical behavior of the Ising model with defects
starting from the disordered initial states with $m_0\ll 1$ also has revealed three stages
of the dynamic evolution. It was shown that the power-law dependences for the
magnetization $m(t)$, the second moment $m^{(2)}(t)$, and the autocorrelation $A(t)$ are
observed in the critical point, which are typical for the pure system in the common interval
[5,60] for observable quantities and for the disordered system in the common interval
[150,800]. In the intermediate time interval the crossover behavior is observed in the dynamic
evolution of the system. The obtained values of exponents demonstrate good agreement within
the limits of statistical errors of simulation and numerical approximations with results
of simulation of the pure Ising model by the short-time dynamics method for the
first time interval and with our results of simulation of the critical relaxation of this
model from the ordered initial state.

\begin{acknowledgments}
This work was supported by the Russian Foundation for Basic Research
through Grants No.~10-02-00507 and No.~10-02-00787 and by Grant No.~MK-3815.2010.2
of Russian Federation President. Our simulations were carried out on the SKIF-MSU in the
Moscow State University and MVS15k in Joint Super Computer Center of Russian Academy of Sciences.
\end{acknowledgments}

\end{document}